\documentclass{PoS}
\usepackage{amsmath,amssymb,graphics}
\usepackage{grffile}
\usepackage{mcite}
\usepackage{graphics,epsfig}
\usepackage{epstopdf}
\graphicspath{{./figs/}}
\usepackage{array}
\newcolumntype{L}[1]{>{\raggedright\let\newline\\\arraybackslash\hspace{0pt}}m{
#1}}
\newcolumntype{C}[1]{>{\centering\let\newline\\\arraybackslash\hspace{0pt}}m{#1}
}
\newcolumntype{R}[1]{>{\raggedleft\let\newline\\\arraybackslash\hspace{0pt}}m{#1
}}

\newcommand{\tr}[0]{\text{tr}}

\title{Confinement From The Gauge Invariant Abelian Decomposition}

\ShortTitle{The Gauge Invariant Abelian Decomposition}

\author{\speaker{Nigel Cundy}\\
         Lattice Gauge Theory Research Center, FPRD, and CTP, Department of Physics \&
    Astronomy,\\ Seoul National University, Seoul, 151-747, South Korea\\
        E-mail: \email{ndcundy@phya.snu.ac.kr}}

\author{Yongmin Cho\\
        Administration Building 310-4, Konkuk University,
Seoul 143-701, Korea}

\author{Weonjong Lee\\
         Lattice Gauge Theory Research Center, FPRD, and CTP, Department of Physics \&
    Astronomy,\\ Seoul National University, Seoul, 151-747, South Korea}

\abstract{A common approach while considering confinement is to study the dominance of an Abelian subgroup of the SU(3) gauge Links. A good way to find the Abelian component of the field is through the Cho-Guan-De gauge invariant Abelian Decomposition, which uses a carefully chosen direction vector $n$ to split the gauge field into an Abelian restricted field and a remnant coloured field. The restricted field can be further subdivided into topological and non-topological terms. We show that there is a choice of $n$ which allows us to exactly represent the Wilson Loop of full QCD as a function of only the restricted Abelian field without requiring any path ordering or additional path integrals. We present numerical evidence showing that the topological part of the restricted field dominates the string tension. We also show that $n$ contains certain topological objects, which, if they exist, will be at least partially responsible for confinement. These leave distinctive patterns in the restricted field strength, and we search for these structures in quenched lattice QCD.}

\FullConference{31st International Symposium on Lattice Field Theory - LATTICE 2013\\
		July 29 - August 3, 2013\\
		Mainz, Germany}

\begin{document}
\section{Introduction}
 We~\cite{Cundy:2012ee,*Cundy:2013xsa} seek to explain the emergence of 
the linear string tension in QCD by studying the Wilson Loop. 
 A common approach, often achieved by fixing to a particular gauge~\cite{Kronfeld1987516,*Suzuki:1989gp,*deForcrand:2000pg,*Stack:1994wm,*Shiba:1994db,*Greensite:2003bk}, is to extract the Abelian part of the 
gauge link (Abelian decomposition), projecting out the coloured, off-diagonal, elements of the gauge link  leaving just an Abelian, colour neutral field which is expected to dominate confinement.
   It is best to respect gauge invariance by using the 
Cho-Duan-Ge (CDG) Abelian 
decomposition~\cite{Cho:1980,*Cho:1981,*Duan:1979,*F-N:98,*Shabanov:1999}.
   The CDG decomposition extracts the components of the gauge field aligned with $N_C-1$ commuting traceless colour vectors $n_j$; our choice of $n_j$ is the main novelty of this study.
   Other recent studies~\cite{Kondo:2005eq,*Kondo:2010pt,*Shibata:2007pi} 
select $n$ (the Abelian theory must be U(1) in that work) through additional dynamical fields which allow the authors to relate the 
string tensions of the Abelian theory and full QCD. 
    Instead, we maximise the Abelian symmetry so that all the possible 
degrees of freedom are included.
   It is possible to choose a specific $n_j$ which 
diagonalises the Wilson Loop and leaves the maximal $U(1)^{N_C-1}$ Abelian 
symmetry for an original SU($N_C$) gauge theory.
   This $U(1)^{N_C-1}$ theory can be studied numerically and modelled theoretically. We believe that certain topological objects contained within the colour fields can provide an explanation of confinement.
  
  In section \ref{sec:2}, we use a particular choice of the CDG decomposition to diagonalise the Wilson Loop, and outline how this might be used to demonstrate a linear static potential. In section \ref{sec:4}, we provide numerical results supporting our analysis, and we conclude in section \ref{sec:5}.

\section{Diagonalisation of Wilson Loops}\label{sec:2}
   A linear static potential, $V(R)$, is a signal for confinement. For a gauge field $A_\mu$, $V(R)$ may be constructed using  $V(R) =- \lim_{T\rightarrow 
\infty} \frac{1}{T}\log \langle \tr\; W[\{R,T\},U] /N_C\rangle$, where $W$ is 
the $R\times T$ Wilson Loop~\cite{wilson:1977}. 
   Consider a Wilson loop of length $L=N\delta \sigma$ parametrised by $\sigma$ around a curve 
$C_s$, a $R\times T$ rectangle in the $x$-$t$ plane,  with $x_\mu(\sigma = 0)= x_\mu(\sigma = L)=s_\mu$, where $P$ represents path ordering,
 \begin{align}
  W[C_s,U] = &\lim_{\delta\sigma \rightarrow 0}\prod_{\sigma = 
0,\delta\sigma,2\delta\sigma,\ldots}^{(N-1)\delta\sigma} U_{\mu(\sigma)}(x(\sigma))&
  U_{\mu}(x) =& P[e^{-ig \int_x^{x + \delta\sigma \hat{\mu}} dx'_\mu 
A_\mu}].
 \end{align}
  We now insert an identity operator $I_\sigma^r = 
\theta^r_\sigma (\theta^r_\sigma)^{-1}$ between each pair of gauge links, with
  $\theta \in U(N_C)$ and $r$ an index identifying the Wilson Loop.
  {We choose $\theta$ so that it diagonalises the gauge links along the 
Wilson Loop.} The index $j$ runs over only the diagonal Gell-Mann matrices.
\begin{align}
  \theta_s^\dagger W[C_s]\theta_s = &\lim_{\delta\sigma \rightarrow 
0}\prod_{\sigma = 0,\delta\sigma,2\delta\sigma,\ldots}^{(N-1)\delta\sigma} 
\theta_\sigma^\dagger U_{\mu(\sigma)}(x(\sigma))\theta_{\sigma+\delta\sigma}\nonumber\\
[\theta_\sigma^\dagger U_{\mu(\sigma)}(x(\sigma)) \theta_{\sigma+\delta\sigma},\lambda^j] = & 
0,\;\;\;\;\;\;\;\;\;\;\;\;\;\;\;\; \lambda^j \text{ diagonal Gell-Mann matrix}, \forall j,U_\mu  
\in C_s.
  \end{align}
    $\theta$ is uniquely defined up to a $U(1)^{N_C}$ transformation $\chi$ (i.e. $\theta \rightarrow \theta\chi$) and 
the ordering of the eigenvectors. By diagonalising the gauge links on all nested Wilson Loops, we can extend this definition of $\theta$ across the entire lattice.
  We now introduce new $SU(N_C)$ fields, $\hat{U}$ and $\hat{X}$, so that
\begin{align}
 {[\theta_\sigma^\dagger \hat{U}_{\mu(\sigma)}(x(\sigma)) 
\theta_{\sigma+\delta\sigma},\lambda^j] =} &{ 0\;\; \forall x,\mu,j};
&U_\mu(x) =& \hat{X}_\mu \hat{U}_\mu\;\; \forall x,\mu;
 &\hat{U}_{\mu}(x) =& U_{\mu}(x)\;\; \forall x,\mu \in C_s.
\end{align}
  This allows us to express the Wilson Loop without any path ordering,
\begin{align}
 \hat{U}_\mu(x) = &\theta_x e^{-i \int_x^{x+\delta\sigma \hat{\mu}} dx'_\mu\hat{u}^j_\mu(x') 
\lambda^j}\theta^\dagger_{x+\delta\sigma \hat{\mu}}; &
\tr W[C_s,U] = & \tr 
W[C_s,\hat{U}] = \tr e^{-ig\oint_{C_s} dx^\mu \lambda^j 
\hat{u}^j_\mu(x)}.
\end{align}
   We can extract the string tension from the Abelian field $\hat{u}$ (a function of $\theta$ and $A_\mu$). We use Stoke's theorem to express the line integral over the Abelian field as a surface integral.
    \begin{align}
  \oint_{C_s} dx^\mu  \hat{u}_{\mu,x}^j =& \int_{x  \in \Sigma,x \not{\in} 
\tilde{\Sigma}} d\Sigma^{\mu\nu} \hat{F}_{\mu\nu}^j   {+ \sum_{n=1}^{\tilde{N}} 
\oint_{\tilde{C}_n} dx^\mu \hat{u}^j_{\mu,x}};&
  \hat{F}_{\mu\nu}^j =& \partial_\mu \hat{u}^j_\nu - \partial_\nu 
\hat{u}^j_\mu.
 \end{align}
   $\Sigma$ represents the planar surface bound by $C_s$.
   $\tilde{\Sigma}_n$ represents the $\tilde{N}$ regions within $\Sigma$ 
(bound by the curves $\tilde{C}_n$ $\in\Sigma$) where $\hat{u}$ is not analytic. $\hat{F}$ and $\hat{u}$ are gauge invariant. Defining $X_0 = 
\frac{1}{2}(\hat{X} +\hat{X}^\dagger)$, 
\begin{multline}
 i\delta\tilde{\sigma} \hat{u}^j_{\mu,x} = \frac{1}{\tr 
(\lambda^j)^2}\text{Im}\left(\;\tr \left[\lambda^j\theta^\dagger_x 
\hat{X}^\dagger_{\mu,x} \theta_x 
\theta_x^\dagger U_{\mu,x}\theta_{x+\delta\tilde{\sigma} 
\hat{\mu}}\right]\right)\\=
\frac{1}{2\tr (\lambda^j)^2}\tr[\lambda^j \theta^\dagger_x 
(\hat{X}_{\mu,x}^\dagger - \hat{X}_{\mu,x}) \theta_x -   2i \lambda^j \delta 
\tilde{\sigma}\theta^\dagger_x [X_0]_{\mu,x} 
gA^a_{\mu,x} \lambda^a\theta_x +  \\ { 2\lambda^j \theta^\dagger [X_0]_{\mu,x} 
\theta_x \delta \tilde{\sigma}\theta_x^\dagger\partial_{\tilde{\sigma}} 
\theta}]+O(\delta\sigma^2).
\end{multline}
  We choose $\hat{X}$ so that 
  {$\tr(\lambda^j\theta^\dagger_x(\hat{X}_\mu(x) - 
\hat{X}_\mu(x)^\dagger)\theta_x) = 0$} and $\tr \hat{X}_\mu(x)$ is maximised.
  If the singularity in $\hat{u}$ occurs over a small region where $A_\mu$ 
and ${X}_0$ are smooth, the $\theta^\dagger \partial_\mu \theta$ term will 
dominate, and
\begin{align}
 \oint_{C_s} dx^\mu  \hat{u}_\mu(x)^j =&   {\sum_n \oint_{\tilde{C}_n} 
d\tilde{\sigma} \tr  [\lambda^j \theta^\dagger [X_0]_{\mu,x} \theta_x \delta 
\tilde{\sigma}\theta_x^\dagger\partial_{\tilde{\sigma}} \theta]} + 
\ldots.\nonumber
\end{align}
  From the above analysis, we see that $\hat{U}$ and $\hat{X}$ are uniquely defined by the equations
\begin{align}
  {\hat{U}_{\mu,x} n_{j,x+\delta\sigma\hat{\mu} }\hat{U}^\dagger_{\mu,x} - 
n_{j,x}=} &  {0} &   {\tr\; n^j (\hat{X} - \hat{X}^\dagger) =} &  {0}&
n_{j,x} \equiv &\theta_x \lambda^j \theta^\dagger_x\nonumber.
\end{align}
  This is a lattice representation of the continuum Cho-Duan-Ge 
gauge-invariant Abelian decomposition~\cite{Cho:1980}, which is known to contain topological 
singularities within the colour field $n$.
  
  Non-analyticities in the $\theta$ field occur when (a) the Wilson Loop has 
degenerate eigenvalues; (b) $A_\mu$ is discontinuous (in the chosen gauge); or 
(c) the situation described below. 
   After gauge fixing, for a SU(2) theory, we parametrise $\theta$ using a complex Givens rotation and a U(1) term,
 \begin{align}
  \theta =& e^{ia\phi} e^{id_3 \lambda^3};& \phi = &\left(\begin{array}{cc} 0 & e^{ic}\\  e^{-ic} 
& 0\end{array}\right);&\overline{\phi} = &\left(\begin{array}{cc} 0 & ie^{ic}\\ -i e^{-ic} 
& 0\end{array}\right).
 \end{align}
 $a$, $c$ and $d_3$ are not 
gauge invariant; $0\le a\le \pi/2$, $c,d_3 \in \mathbb{R}$. In SU(3), we construct $\theta$ from three Givens terms and a $U(1)\times U(1)$ matrix parametrised by $d_3$ and $d_8$.
The arbitrary parameters $d_3$ and $d_8$ do not affect the  field $n$ 
(they can be chosen to be zero). 
  {$c$ is ill defined at $a = 0$ or $a = \pi/2$.} We 
parametrise space-time around one of these points as
\begin{gather}
{ (t,x,y,z) \equiv r 
(\cos\psi_3,\sin\psi_3\cos\psi_2,\sin\psi_3\sin\psi_2\cos\psi_1,
\sin\psi_3\sin\psi_2\sin\psi_1),}\label{eq:72}
\end{gather}
with $r=0$ at $a = \pi/2$. In SU(2), by writing $c = \nu_n \psi_3$ for an integer 
gauge-invariant winding number $\nu_n\neq 0$, and using {$
\theta^\dagger\partial_\sigma \theta = e^{-id_3\lambda^3}\big[ i \partial_\sigma 
a \phi + i \lambda^3 \partial_\sigma d +
 i \sin a \cos a \overline{\phi}\partial_\sigma c - i \sin^2 a \partial_\sigma c 
\lambda^3 \big]e^{id_3 \lambda^3}
$},
  we may integrate around a curve at fixed $a = a_{0n}$ surrounding the
structure in $\hat{F}$ to obtain 
\begin{gather}
 \oint_{C_s} dx^\mu  \hat{u}_\mu(x)^j =   {\sum_{n=1}^{\tilde{N}} 2\pi \nu_n 
\sin^2 a_{0n}\tr  [ [X_0]_{\mu,x} ]} + \ldots.
\end{gather}
  If the number of structures, $\tilde{N}$, within the Wilson loop is 
proportional to the area enclosed by the loop, as might be expected, then this leads to an area law 
string tension and confinement.

   We parametrise $a = \frac{\pi}{2} - G(r,\psi_1,\psi_2,\psi_3)$ and $c = 
J(\psi_3)$ for unknown gauge-dependent functions $G$ and $J$. Then, we can 
calculate the topological ($\theta$) contribution $H^3_{\mu\nu}$ to the field 
strength $\hat{F}^3_{\mu\nu}$. $H_{\mu\nu}^j = \frac{1}{8g}\tr n_j[\partial_\mu 
n_k,\partial_\nu n_k]$. 
   In SU(2), with $G\sim \partial_i G \equiv \frac{\partial G}{\partial \psi_i} \sim r^\xi$; $\partial_r G \sim r^{\xi-1}$ and $\xi > 0$
 \begin{align}
 B_y=&\frac{1}{g}\sin 2G \bigg(\underbrace{\partial_1 G \partial_3 J  
\frac{yxt}{r^2 r_{xyz} r_{yz}^2}}_{\text{t-string}} + \underbrace{\partial_2 G 
\partial_3 J  \frac{zt}{r^2 r_{xyz} r_{yz}}}_{\text{t-string}}\bigg)
&
 B_x =&\frac{1}{g}\sin 2G \bigg( \underbrace{\partial_1G  \partial_3 J 
\frac{t}{r_{xyz} r^2}}_{\text{t-string}} \bigg)\nonumber\\
E_x= &-\frac{1}{g}\sin2G\bigg(\underbrace{\partial_rG  \partial_3 J \frac{x}{r 
r_{xyz}}}_{\text{point}} - \underbrace{\partial_2 G \partial_3 J 
\frac{r_{yz}}{r^2 r_{xyz}}}_{\text{point}}  \bigg) & B_z=&-\frac{1}{g}\sin 2G \bigg( \underbrace{\partial_1 G \partial_3 J 
\frac{zxt}{r^2 r_{yz}^2 r_{xyz}}}_{\text{t-string}}\bigg)
\nonumber
\end{align}\vspace{-0.8cm}
\begin{align}
 E_y=&-\frac{1}{g}\sin 2G \bigg(\underbrace{\partial_r G \partial_3 J  
\frac{y}{r_{xyz} r}}_{\text{point}} - \underbrace{\partial_1 G \partial_3 J 
\frac{z r_{xyz}}{r^2 r_{yz}^2}}_{\text{x-string}} + \underbrace{\partial_2 G 
\partial_3 J\frac{xy}{r^2 r_{xyz}r_{yz}}}_{\text{point}} \bigg)\nonumber\\
 E_z=&-\frac{1}{g}\sin 2G\bigg(\underbrace{\partial_r  G\partial_3 J \frac{z}{r 
r_{xyz}}}_{\text{point}} - \underbrace{\partial_1 G \partial_3 J \frac{y 
r_{xyz}}{r^2 r_{yz}^2}}_{\text{x-string}} 
-\underbrace{\partial_2G\partial_3J\frac{zx}{r_{yz}r_{xyz} r^2}}_{\text{point}}  
\bigg) 
.
\end{align}
  a $\mu$-string is a 1-Dimensional object parallel to the $\mu$-axis; a 
point is a structure where the maximum falls at least as $1/r$ in all directions (for $\xi = 1$).
  After rotating the coordinate system 
consistent with the overall symmetry, we find the following structures 
in the electromagnetic field strength:
\begin{center}
\begin{tabular}{l l l l l l l}
\hline
Parametrisation&$E_x$&$E_y$&$E_z$&$B_x$& $B_y$&$B_z$
\\
\hline
Equation (\ref{eq:72})&point&x-string&x-string&t-string&t-string&t-string\\
$t\leftrightarrow x$&point & x-string&x-string&x-string&t-string&t-string\\
$y\leftrightarrow z$&point&x-string&x-string&t-string&t-string&t-string\\
$t\leftrightarrow x$,$y\leftrightarrow 
z$&point&x-string&x-string&x-string&t-string&t-string\\
\hline\hline
\end{tabular}
\end{center}
  If these topological structures exist, they will reveal themselves as 
points in the $xt$ component of the field strength, and either points, 
$x$-strings or $t$-strings in the other components of the field strength.

\section{Numerical Results}   \label{sec:4}

\begin{figure}
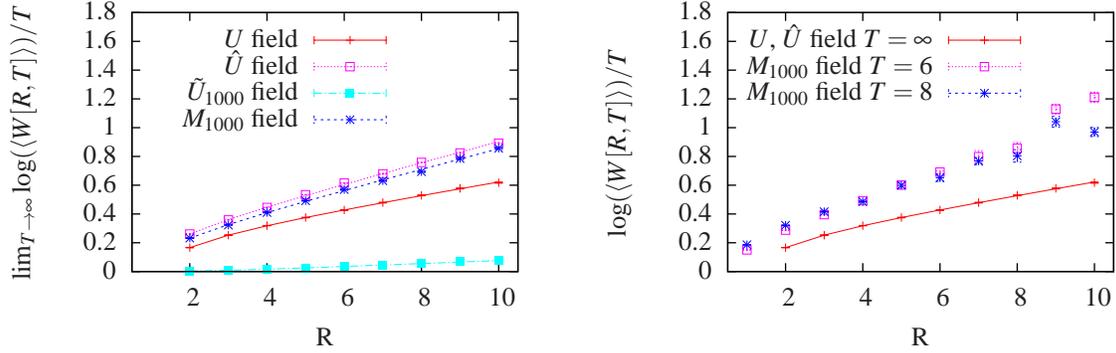

\begin{center}
\begin{tabular}{cc}
{\small
\input{figs/b8.52p.tex}
}&
{\small
\input{figs/b8.52realp.tex}
}
\end{tabular}
\end{center}\vspace{-0.8cm}
\caption{The string tension calculated from (R,T) Wilson Loops on the $\beta = 8.52$ ensemble, for $\theta$ 
is fixed for each configuration (left) and early results with $\theta$ recalculated for each Wilson Loop (right). Our data where $\theta$ is recalculated is not yet good enough for a reliable extrapolation to $T=\infty$.}\label{fig:1}
\end{figure}

We used quenched Luscher-Weisz~\cite{TILW,*TILW2,*TILW4} lattice QCD configurations. Our Lattice spacings and lattice volumes are shown in table \ref{tab:1}. To preserve gauge invariance, we use a stout smeared gauge field 
$\tilde{U}_p$ (after a large number, $p$, of smearing steps) during our construction of the topological field,
  $M$, taken from the Abelian decomposition of $\tilde{M}_p = 
\theta^\dagger \tilde{U}_p \theta$; $\tilde{U}$ will not contribute to 
the string tension. In figure \ref{fig:1} and table \ref{tab:2} we show results for the string tension for the gauge fields $U$, 
$\hat{U}$, and $M$, the topological ($\theta$) part of $\hat{U}$. To save computer time, initial results used a single $\theta$ for each Wilson Loop on a configuration, breaking the identity between $\tr W[C_s,U]$ and $\tr W[C_s,\hat{U}]$. The $\hat{U}$ and $M$ string tensions (seen in the slope of the curves) are in good agreement.
   

\begin{table}
\begin{center}
\begin{tabular}{l| l l l l l}
\hline
Name&Lattice size & L (fm)&$\beta$&$a$ (fm)& $\#$\\
\hline
8.0&$16^3 \times 32$& 2.30& 8.0&0.144(2) &91
\\
8.3&$16^3 \times 32$&1.84& 8.3 &0.114(1) &91
\\
8.52&$16^3 \times 32$&1.58 & 8.52 &0.099(1) &82
\\
8.3L&$20^3 \times 40$&2.30 & 8.3&0.112(5) &54\\
\hline\hline
\end{tabular}
\end{center}\vspace{-0.5cm}
\caption{Parameters for the ensembles.
$\#$ is the number of configurations. $L$ the 
physical spatial extent.}\label{tab:1}
\end{table}

\begin{table} 
\begin{center}
\begin{tabular}{l|llllllll}
\hline
&$U$&$\hat{U}$&$\tilde{{U}}_{600}$&$M_{600}$&$\tilde{{U}}_{800}$&$M_{800}$&$\tilde{{U}}_{1000}$&$M_{1000}$ \\
\hline
8.0&0.094(2)&0.116(4)&0.0273(2)&0.103(10)&0.0213(1)&0.104(8)&0.0174(1)&0.105(9)
\\
8.3&0.0590(8)&0.095(2)&0.0185(1)&0.087(5)&0.0147(1)&0.087(5)&0.0122(1)&0.087(5)
\\
8.52&0.0442(6)&0.077(1)&0.0149(2)&0.076(3)&0.0124(2)&0.076(3)&0.0106(2)&0.077(3)
\\
8.3L&0.057(5)&0.099(1)&0.0179(1)&0.099(2)&0.0144(1)&0.099(2)&0.0121(1)&0.098(2)
\\
\hline\hline
\end{tabular}
\end{center}\vspace{-0.5cm}
\caption{The string tension for the ensembles ($\theta$ fixed/configuration)} \label{tab:2}

\end{table}

\begin{figure}
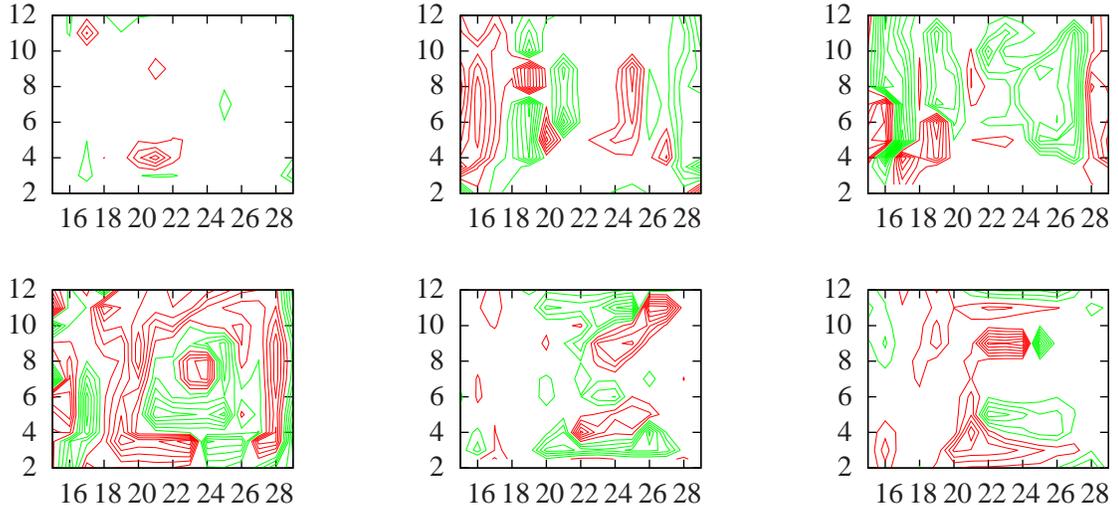

 \begin{center}

\begin{tabular}{ccc}
\input{figs/E_x_b8-52_70_xtp.tex}&\input{figs/E_y_b8-52_70_xtp.tex}&\input{
figs/E_z_b8-52_70_xtp.tex}\\
\input{figs/B_x_b8-52_70_xtp.tex}&
\input{figs/B_y_b8-52_70_xtp.tex}&
\input{figs/B_z_b8-52_70_xtp.tex}
\end{tabular}
\end{center}  \vspace{-0.8cm}
  \caption{ Contour plots for the field strength for the $x$ (left), $y$ 
(middle) and $z$ (right) components of the restricted electric (top) and 
magnetic (bottom) fields on an X (y-axis)-T (x-axis) planar slice of the lattice.  Red indicates positive field strength, green negative field strength.}\label{fig:2} 
\end{figure}
\begin{figure}
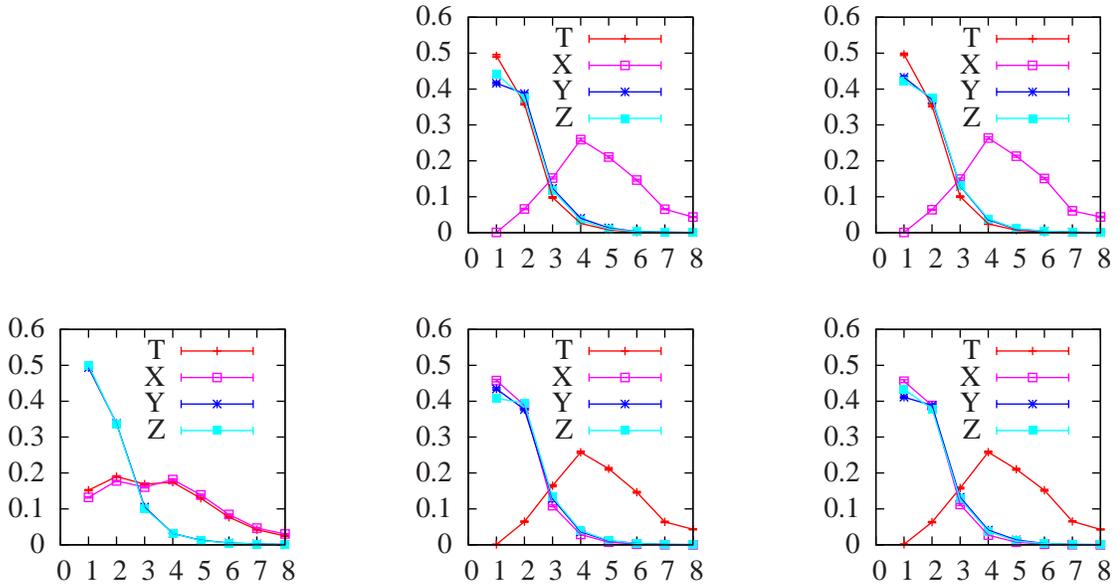

 \begin{center}
  \begin{tabular}{ccc}
& \input{figs/Cluster_extent_E_y_b8_52p.tex}& 
\input{figs/Cluster_extent_E_z_b8_52p.tex}\\
\input{figs/Cluster_extent_B_x_b8_52p.tex} 
&\input{figs/Cluster_extent_B_y_b8_52p.tex} 
&\input{figs/Cluster_extent_B_z_b8_52p.tex} 
     \end{tabular}
 \end{center}\vspace{-0.8cm}
\caption{The spatial extent of the clusters containing more than four lattice sites along the four spatial directions for the $x$ (left) $y$ (middle) and $z$ (right) components of the electric (top) and magnetic (bottom) fields. The $X$ axis gives the length of the cluster in a given direction; the $Y$ axis the proportion of clusters with that length.}\label{fig:4}
\end{figure}

  Figure \ref{fig:2} displays contour plots showing the distribution of the various components of the field strength.  The dominant structures appear to be points or lines in the expected directions.
   This is confirmed by a cluster analysis. We identify clusters as sign-coherent regions of field strength with $|\hat{F}_{\mu\nu}| >1$ for each $\mu,\nu$. We then compare the size, shape and orientation of the clusters with the model expectations of the topological objects in the field strength.
\begin{figure}
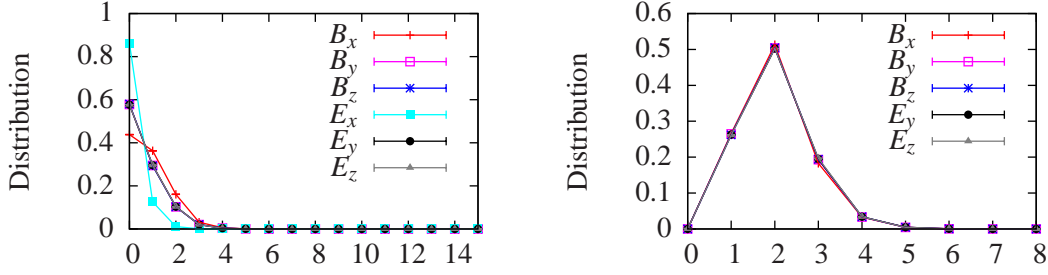

\begin{center}
 \begin{tabular}{cc}
 \input{figs/Cluster_neighbours_b8_52p.tex}& \input{figs/Cluster_f_neighbours_b8_52p.tex}
 \end{tabular}
\end{center}\vspace{-0.8cm}
\caption{ The average number of nearest neighbours within a cluster for each lattice site 
in the cluster (left) and the same analysis only including points within clusters extended over at least four lattice sites (right).}\label{fig:3}
\end{figure}
In figures \ref{fig:3} and \ref{fig:4}, we investigate whether the objects of within the Abelian Field strength have the shapes expected from the theory. Figure \ref{fig:3} investigates the dimensionality of the clusters by investigating the number of nearest neighbours of each site in the cluster. Excluding the smallest clusters, the majority of lattice sites have two neighbouring sites within the same cluster, suggesting that these objects are one-dimensional. In figure \ref{fig:4} we investigate the orientations of these strings (excluding the smallest point-like structures from the analysis, including all structures in $E_x$). As expected, $E_y$ and $E_z$ are extended along the $x$-axis, $B_y$ and $B_z$ along the $t$-axis, with $B_x$ extended along 
both the $x$ and $t$ axes.

  \section{Conclusions}   \label{sec:5}
     We have suggested that, by introducing a carefully tuned field, 
$\theta$, it is possible to diagonalise the gauge links within a Wilson Loop 
without introducing additional path integrals or dynamical variables, giving a $U(1)^{N_C-1}$ 
Abelian theory (a CDG decomposition) which can be used to calculate the string tension.
     This theory can be studied numerically, and modelled theoretically.
     As expected, the coloured fields do not contribute to 
confinement. There may be certain topological singularities within $\theta$ which 
contribute to the string tension, giving characteristic structures appearing in the Abelian 
field strength tensor.
     We have confirmed numerically that the topological term accounts for 
all of the string tension, and that the structures within the field strength 
have the same dimensionality and directions as expected from the model.
\section*{Acknowledgements}
Numerical calculations used servers at Seoul National University. Funding was provided by the BK21 program of the NRF, Republic of Korea. The research of W.~Lee is supported by the Creative Research
Initiatives Program (2012-0000241) of the NRF grant funded by the
Korean government (MEST).
W.~Lee would like to acknowledge the support from KISTI supercomputing
center through the strategic support program for the supercomputing
application research [No. KSC-2011-G2-06].
 YMC is supported in part by NRF grant (2012-002-134)
funded by MEST.

 \bibliographystyle{JHEP_mcite.bst}

\bibliography{weyl}

\end{document}